\documentclass{article}

\usepackage{arxiv}

\usepackage[utf8]{inputenc} 
\usepackage[T1]{fontenc}    
\usepackage{hyperref}       
\usepackage{url}            
\usepackage{booktabs}       
\usepackage{amsfonts}       
\usepackage{nicefrac}       
\usepackage{microtype}      
\usepackage{lipsum}		
\usepackage{subcaption}
\usepackage{graphicx}
\usepackage{tikz}
\usetikzlibrary{arrows}
\usepackage{amsthm}
\usepackage{commath}
\usepackage{float}

\theoremstyle{definition}
\newtheorem{definition}{Definition}

\title{Quantitative analysis of cryptocurrencies transaction graph}

\author{
	Amir Pasha Motamed\\
	School of Electrical and Computer Engineering\\
	College of Engineering\\
	University of Tehran\\
	\texttt{a.motamed@ut.ac.ir} \\	
	\And
	Behnam Bahrak\\
	School of Electrical and Computer Engineering\\
	College of Engineering\\
	University of Tehran\\
	\texttt{bahrak@ut.ac.ir} \\
}

\begin{document}
\maketitle

\begin{abstract}
Cryptocurrencies as a new way of transferring assets and securing financial transactions have gained popularity in recent years. Transactions in cryptocurrencies are publicly available, hence, statistical studies on different aspects of these currencies are possible. However, previous statistical analysis on cryptocurrencies transactions have been very limited and mostly devoted to Bitcoin, with no comprehensive comparison between these currencies. In this study, we intend to compare the transaction graph of Bitcoin, Ethereum, Litecoin, Dash, and Z-Cash, with respect to the dynamics of their transaction graphs over time, and discuss their properties. In particular, we observed that the growth rate of the nodes and edges of the transaction graphs, and the density of these graphs, are closely related to the price of these currencies. We also found that the transaction graph of these currencies is non-assortative, and the degree sequence of their transaction graph follows the power law distribution.
\end{abstract}


\section{Introduction}
Cryptocurrencies have made it possible for a financial system to perform transactions without the need for a centralized authority while keeping the transaction details and money generation clear and publicly available. Despite this transparency, people's identities are hidden, and they can transact anonymously. All transaction information of a cryptocurrency is usually stored in a distributed public ledger, named blockchain. The tasks of recording, updating, and maintaining the blockchain is the responsibility of network users for each coin, whose identities are unknown, and rewards have been created to provide them with sufficient incentives to do so, making the network up and running. Although the system is running by anonymous people, committing fraud and transaction alteration is computationally impossible. This level of security is guaranteed by cryptographic algorithms, and as long as these algorithms are secure, cryptocurrencies integrity is protected.
\par
Bitcoin is the first cryptocurrency created by an anonymous person or group of people by the nickname Satoshi Nakamoto, which established a decentralized money transfer system using blockchain \cite{nakamoto2008bitcoin}. Subsequently, other cryptocurrencies, which are usually referred to as altcoins, were created by adding more capabilities and offering alternative design criteria. Ethereum was introduced by Vitalik Buterin in 2015 and is the first blockchain-based distributed computing platform to consider the concept of executable smart contracts \cite{buterin2014next}. It is one of the most influential and widely-used cryptocurrencies introduced after Bitcoin. Litecoin is also one of the earliest cryptocurrencies that is technically very similar to Bitcoin and has only slight differences with it \cite{litecoinwiki}. For example, Litecoin uses the Scrypt hash function instead of SHA256 for proof of work, and records transactions in the blockchain four times faster than Bitcoin. Litecoin is created as a hard fork of Bitcoin, and has a separate blockchain. Dash is another cryptocurrency which is quite similar to Bitcoin and uses the X11 hashing algorithm for proof of work \cite{dashwhitepaper}. Similar to Litecoin, Dash has a separate blockchain, with transactions speed 4 times faster than Bitcoin. Z-Cash is a highly secure cryptocurrency that uses zero-knowledge proofs, as a result of which privacy and anonymity of users is significantly enhanced \cite{hopwood2016zcash}. 
\par
In all of the mentioned cryptocurrencies, the ability to transfer money is the basic and common core capability. Using the blockchain data of each of these currencies, the transactions in which they occur can be accessed. As a result, it is possible to analyze transactions in these currencies from different aspects and perform a variety of statistical analyses on them. In particular, it is possible to examine a real network of financial transactions for each cryptocurrency.
\par
In this paper, the financial exchange network of the five aforementioned cryptocurrencies has been studied and several statistical metrics and network measures are calculated, and their meanings are discussed. From a perspective, this financial exchange network can be seen as a social network. In social networks, nodes are individuals, and the edges between them can be friendships or other social relationships. In the transaction graph of a cryptocurrency, vertices are accounts (or addresses) in the currency network, and the edges between them are transactions between those accounts. Since these accounts have hidden identities, they do not represent the true identities of individuals. Note that a person can create multiple accounts, and it is almost impossible to link these accounts, and detect that they belong to the same individual. There are graph analytics methods and heuristics to link some of the accounts \cite{nick2015data}, but since these techniques are prone to errors and cannot detect all related accounts, we do not use any of these methods for linking accounts and merging their corresponding nodes in the transaction graph.
Our contributions can be summarized as follows:
\begin{enumerate}
	\item
	We compare the structural properties of the transaction graphs of five widely-used cryptocurrencies.
	\item
	We discuss the relation between the transaction graph properties with technical aspects and historical events of each coin.
	\item
	We investigate the evolution of the transaction graph over time and study the effect of supply and demand, and price of each coin on the transaction graph.
\end{enumerate}

\section{Related work}
Various studies have been conducted on cryptocurrency transaction networks from different perspectives. Among these studies, there is no comprehensive review, and most of them have focused on one or two specific coins, especially Bitcoin and Ethereum, and used outdated blockchain data which does not cover recent developments in the field. In most of these studies the transaction graph is investigated statically and its dynamics and evolution over time are not considered.
\par
In a study by Chen et al., the relationships between smart contracts and people's accounts and money flows in the Ethereum network is investigated \cite{chen2018understanding}. Ron and Shamir in \cite{ron2013quantitative}, analyzed the bitcoin transaction graph statically. In another study on Bitcoin, Maesa et al. have applied a heuristic for deanonymization and extracted user's graph by merging addresses that belong to the same individuals. They analyzed the distances between nodes and studied graph metrics such as density and phenomena like the rich-get-richer phenomenon \cite{maesa2018data}. But these studies are only limited to Bitcoin, and with modern wallets and the advent of mixers \cite{mixerbitcoin}, the deanonymization heuristic has become ineffective. In another related word by Fleder et al., the authors have studied the relationship between social networks and bitcoin transactions, using posts in online forums and social networks to find out the actual identity of people in Bitcoin \cite{fleder2015bitcoin}. Their analysis and calculations on the transaction graph is limited, and their deanonymization heuristic is no longer valid due to the existence of mixers. Liang et al. in \cite{liang2018evolutionary}, have compared the transaction graphs of Bitcoin, Ethereum and Namecoin. Currently, Namecoin is no longer active and it is not in the list of top 100 coins according to their market capitalization \cite{cap2018cryptocurrency}. Kondor et al. also conducted a study on the Bitcoin transaction graph, which examined how the Bitcoin transaction graph evolved, but their study is limited to Bitcoin \cite{kondor2014rich}. In another study by Guo et al., the Ethereum transaction graph is investigated. However, this study is limited to the Ethereum and only deals with a small part of the blockchain at two specific time spots \cite{guo2019graph}.
\section{Methodology}
\subsection{Dataset}
The data used in this study were obtained directly from the blockchain of the cryptocurrencies. There are several ways to get these information, and we used two different methods for data collection. For Bitcoin, Ethereum, Litecoin, and Dash, we obtained blockchain data from their peer-to-peer network using their client software. These data are stored in binary format and needed to be converted into human-readable formats such as comma-separated values (CSV) for further analysis. These binary data can be converted by parsers which output several large CSV files. These files contain each transaction details including the timestamp, the number of inputs and outputs of the transaction, the incoming and outgoing addresses, and other related information which is stored in the blockchain. We used \texttt{Ethereum ETL} for Ethereum \cite{ethereumetl2019}, \texttt{Rusty Blockparser} for Litecoin \cite{rustyblockparser2019}\, a modified version of bitcoin core for Bitcoin cite{bitcoinparser2019}. We also made a custom parser for parsing Dash blockchain. To build the transaction graph, we need database operations like Join and Select. Due to the high volume of data, we used \texttt{Apache Spark}, which is one of the most well-known big data processing tools, to perform these operations \cite{spark2019}. It has a programming interface called \texttt{PySpark} for Python language that can optimize SQL statements runs on bulk data \cite{pyspark2019}. To obtain Z-Cash blocks, we used JSON API of SoChain \cite{sochain2019cryptocurrency}, an online blockchain explorer, which provided the information of all blocks in JSON format and then the transaction graph was constructed \cite{sochain2019cryptocurrency}.
\subsection{Building transaction graph}
The blockchain structure for each of these cryptocurrencies is different, but some are very similar. For example, the Bitcoin and Litecoin blockchains are very similar, but the Ethereum blockchain has a completely different structure because of its nature and sophistication. But in all of them, the transaction information is contained within the blocks. In each block, a certain number of transactions can be placed. In general, blockchains can be divided into two categories: UTXO (Unspent transaction output)-based and account-based. In the UTXO-based blockchains, each transaction input is linked to an output of a previous transaction. In other words, current transactions in a block are spending the outputs of previous transactions and generating new outputs to be spent in subsequent transactions. In the outputs of the transactions, the addresses to which the output values belong are placed. But in the account-based blockchains, the addresses of the incoming and outgoing accounts are explicitly stated. The blockchain of Bitcoin, Litecoin, Z-Cash, and Dash use UTXO-based output types, and each transaction contains several inputs and outputs. But in Ethereum, whose blockchain is account-based, each transaction has only one input and one output. In a transaction graph, nodes are accounts addresses, and the edges are the transactions between these accounts. In this study, we considere a transaction graph as an unweighted undirected graph, but in some analyses, we use a directed version of that graph. Given that each block of the blockchain has a timestamp, we have divided the timeline into monthly intervals and created a transaction graph for each month that only includes the transactions in the blocks of that month. To make a transaction graph from a set of transactions, we place one edge from each input address and to each output address in transactions in the graph. For Coinbase transactions, that include the block generation reward given to the miners and the inputs do not refer to a previous transaction outputs, we considered a supernode as its input and one edge of that supernode to each miner address.
\begin{figure}
	\centering
	\begin{subfigure}[t]{0.46\linewidth}
		
		\centering
		\begin{tabular}{ | c | c |}
			\hline
			\multicolumn{2}{|c|}{Transaction} \\
			\hline
			Inputs & Outputs  \\
			\hline
			A & D \\ 
			\hline
			B & E \\
			\hline
			C & \\
			\hline
		\end{tabular}\\
		\vspace{1.48cm}
		\caption{An example of a transaction with 3 Inputs and 2 Outputs}
		\label{fig:exampletransactiondetail}
	\end{subfigure}
	\hspace{0.2cm}
	\begin{subfigure}[t]{0.46\linewidth}
		\centering
		\begin{tikzpicture}[baseline=(current bounding box.center),>=stealth',shorten >=1pt,auto,node distance=3cm,
		thick,main node/.style={circle,draw,font=\sffamily\bfseries}]
		
		\node[main node] (1) at (0,4) {A};
		\node[main node] (2) at (0,2) {B};
		\node[main node] (3) at (0,0){C};
		\node[main node] (4) at (2,3) {D};
		\node[main node] (5) at (2,1) {E};
		
		\path[every node/.style={font=\sffamily\small}]
		(1) edge node [left] {} (4)
		edge node [left] {} (5)
		(2) edge node [left] {} (4)
		edge node [left] {} (5)
		(3) edge node [left] {} (4)
		edge node [left] {} (5);
		\end{tikzpicture}
		\caption{Transaction graph based on the transaction in Fig. \ref{fig:exampletransactiondetail}}
		\label{fig:examplegraph}
	\end{subfigure}
	\caption{Generation of a transaction graph}
	\label{fig:exampletransaction}
\end{figure}
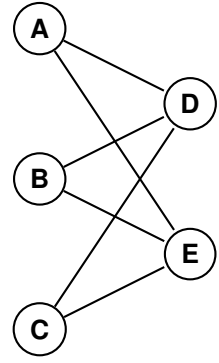
\par
\begin{definition}
	\textit{Monthly Transaction Graph (MTG)} for each coin is a graph that is represented in the form $MTG_n=(V_n,E_n)$ where $V_n$ is the set of nodes of this graph and each node $v \in V_n$ is an account address and $V_n$ is the set of all account addresses that appeared in the $(n+1)$-th month since beginning of the activity of that coin. $E_n$ is the set of edges of this graph where each edge $e=\{v^{\prime},v^{\prime\prime}\}$ shows transfer between two vertices $v^{\prime},v^{\prime\prime} \in V_n$. For example, in Fig. \ref{fig:exampletransaction}, a transaction with three inputs and two outputs is illustrated. 
\end{definition}

\begin{definition}
	\textit{Cumulative Monthly Transactions Graph (CMTG)} for each coin is a graph that is represented in the form $CMTG_n=(V_{C_n},E_{C_n})$ where $V_{C_n}$ and $E_{C_n}$ are the set of its nodes, and its edges, respectively.\\
	$V_{C_n}$ is equal to:
	\begin{equation}
	\centering
	V_{C_n} = \bigcup_{i=0}^n V_i
	\end{equation}
	where $V_i$ is the set of nodes of $MTG_i$.\\
	Similarly $E_{C_n}$ is defined as:
	\begin{equation}
	E_{C_n} = \bigcup_{i=0}^n E_i
	\end{equation}
	where $E_i$ is the set of edges of $MTG_i$. In other words, $CMTG_n$ is the union of the graphs $MTG_0$, $MTG_1$, $MTG_2$, $\cdots$ , $MTG_n$, and is containing the addresses and transactions between them since the creation of that coin until end of $(n+1)$-th month.
\end{definition}

\subsection{Measurements}
There are various metrics for quantitative comparison between transaction graphs of different cryptocurrencies. As mentioned earlier, the transaction graph can be viewed as a social network graph, and all metrics that can be calculated on social networks can also be studied for the transaction graph. We use the most common metrics that are meaningful in the context of transaction graphs and has a relation with technical details and historical events in the timeline of each coin. Given the large size of these gigantic graphs, we only investigate the metrics that might be calculated in an acceptable period of time. In what follows, we introduce the metrics that are calculated on the cryptocurrencies transaction graph in this study. 

\begin{definition}
	\textit{Clustering coefficient} shows the tendency of graph vertices to create a cluster with other vertices in the graph, and is defined as:
	\begin{equation}
	C = \frac{3 \times \textrm{number of triangles}}{\textrm{number of triads}}
	\end{equation}
	A triad is a set of 3 nodes that at least two pairs of them are connected. A triangle is a set of 3 nodes that all three pairs of them are connected and each triangle is consisting of 3 triads. Since calculating the exact value of the clustering coefficient is hard in large graphs, we used an approximate method. In this method, instead of counting all triangles and triads, we randomly selected a specific number of triads and check that is it triangle or not. The estimated clustering coefficient was the percentage of triads that they were also a triangle. As the number of randomly selected triads increases, the estimated clustering coefficient will be closer to the exact value.
\end{definition}
\begin{definition}
	\textit{Density} of an undirected graph with node set $V$ and edge set $E$ is defined as:
	\begin{equation}
	D = \frac{2 \times n(E)}{n(V) \times (n(V)-1)}
	\end{equation}
	where $n(E)$ is the number of graph edges and $n(V)$ is the number of its vertices.
\end{definition}
\begin{definition}
	\textit{Edge-to-vertex ratio} is calculated by dividing the number of edges of a graph to the number of its vertices.
\end{definition}
\begin{definition}
	\textit{Size of maximum clique} in a graph is the number of vertices of its largest complete subgraph. A complete subgraph is a set of nodes and edges in a graph where every pair of nodes is connected by an edge.
\end{definition}
\begin{definition}
	The \textit{assortativity coefficient} of a graph indicates the tendency of the graph vertices to attach to other vertices that are similar to them. The similarity of two nodes is usually measured by their degrees, and the assortativity coefficient is calculated by the Pearson correlation coefficient of degree between pairs of linked nodes.
	
	This metric has a numeric value between $-1$ and $1$. The value of $1$ indicates that the graph is perfectly assortative and the vertices tend to have an edge with other vertices of similar degree. A value of $-1$ indicates that the graph is completely disassortative and its vertices tend to link to vertices with different degrees. An assortativity of $0$ shows that the graph is non-assortative and its vertices are neutral and do not exhibit a tendency for a particular type of vertices.
\end{definition}
\begin{definition}
	The \textit{repetition ratio} in the transaction graph indicates the percentage of repetitive nodes or edges in the MTG in a month compared to the previous month. The repetition ratio of the nodes in the $n+1$-th month is defined as follows:
	\begin{equation}
	RR_{V_i} = \frac{n(V_{MTG_i} \cap V_{MTG_{i-1}})}{n(V_{MTG_i})}
	\end{equation}
	where $V_{MTG_i}$ is the size of node set of graph $MTG_i$.\\
	The repetition ratio of the edges in the $n+1$-th month is similarly defined as:
	\begin{equation}
	RR_{E_i} = \frac{n(E_{MTG_i} \cap E_{MTG_{i-1}})}{n(E_{MTG_i})}
	\end{equation}
	where $E_{MTG_i}$ represents the number of edges of graph $MTG_i$.\\
	Since the repetition ratio for each month is calculated using to the previous month MTG, the repetition ratio for the first month is not defined.
\end{definition}
\section{Results}
In this section we present the results of our comprehensive investigation on cryptocurrencies transaction graph. Because of the large volume of the extracted graphs, we need high computational power and memory to store them and perform calculations on them. We performed our analyses and calculations on a server with eight cores of Intel(R) Xeon(R) CPU E5-2630 v4 $@$ 2.20GHz processor along with 3TB of disk storage and 80GB of RAM.
\begin{figure}
	\centering
	\begin{subfigure}[t]{0.49\linewidth}
		\includegraphics[width = 1\linewidth]{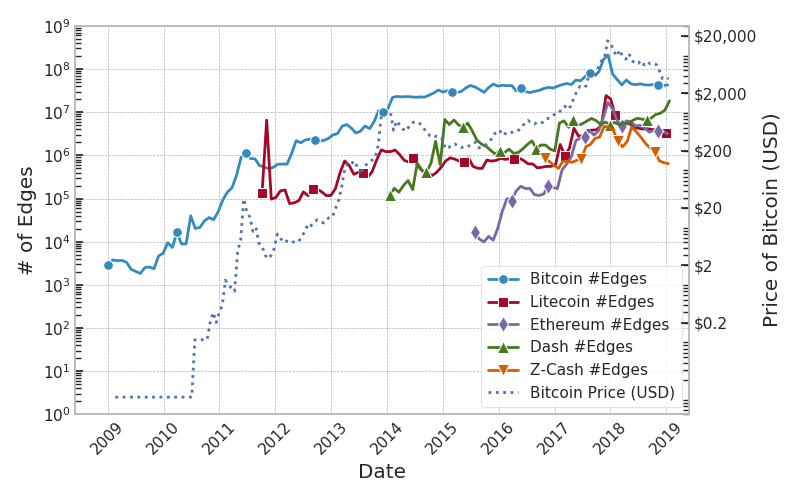}
		\caption{Number of edges}
		\label{fig:graphsizeofedges}
	\end{subfigure}
	\begin{subfigure}[t]{0.49\linewidth}
		\includegraphics[width = 1\linewidth]{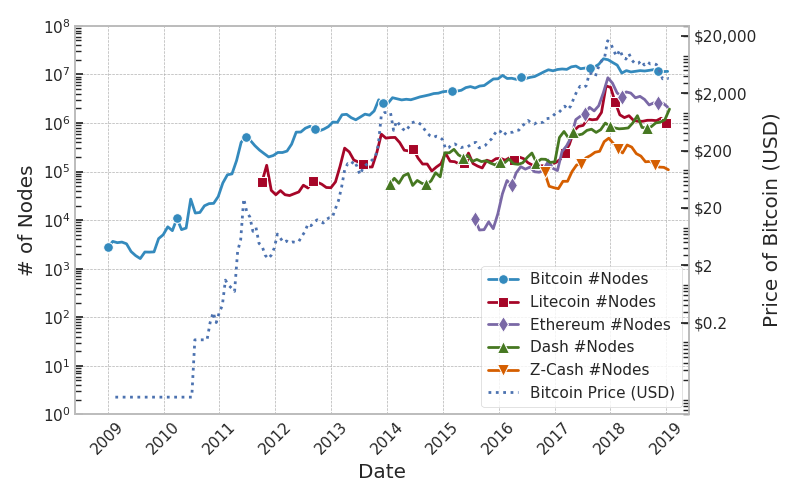}
		\caption{Number of nodes}
		\label{fig:graphsizeofnodes}
	\end{subfigure}
	\caption{Size of MTG graph over time}
\end{figure}
\begin{figure}
	\centering
	\begin{subfigure}[t]{0.49\linewidth}
		\includegraphics[width = 1\linewidth]{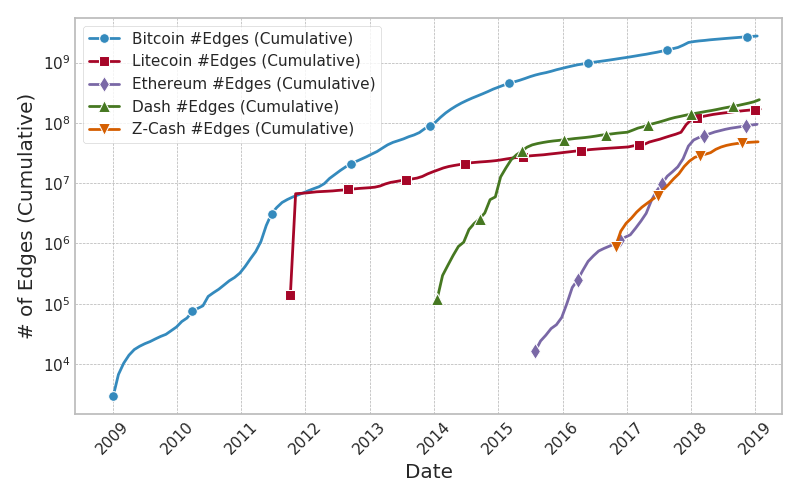}
		\caption{Number of edges}
		\label{fig:graphsizeofedgescumulative}
	\end{subfigure}
	\begin{subfigure}[t]{0.49\linewidth}
		\includegraphics[width = 1\linewidth]{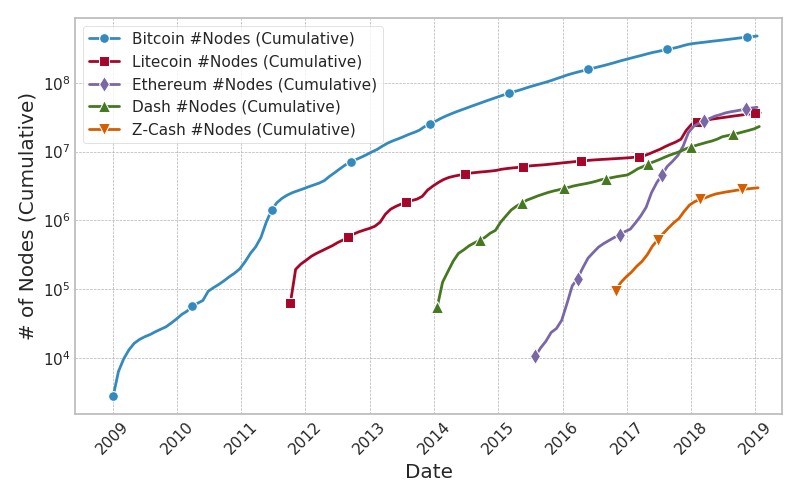}
		\caption{Number of nodes}
		\label{fig:graphsizeofnodescumulative}
	\end{subfigure}
	\caption{Size of CMTG graph over time}
\end{figure}
\begin{table}
	\centering
	\caption{Correlation of price with size of MTG graph}
	\begin{tabular}{| c | c | c |}
		\cline{2-3}
		\multicolumn{1}{c|}{} & with \# of edges & with \# of nodes \\
		\hline
		Bitcoin & $0.7898$ & $0.7652$ \\ 
		\hline
		Litecoin & $0.9079$ & $0.9133$ \\
		\hline
		Ethereum & $0.6674$ & $0.6991$ \\
		\hline
		Dash & $0.3485$ & $0.5642$ \\ 
		\hline
		Z-Cash & $0.2177$ & $0.2443$ \\
		\hline
	\end{tabular}
	\label{tab:pricecorrelation}
\end{table}
\par
The first study addresses the time series of the size of MTG and CMTG graphs. Since the start date of each coin is different, each coin’s time series is plotted from the date of its creation (genesis block creation date). Fig. \ref{fig:graphsizeofedges} and \ref{fig:graphsizeofnodes} illustrate the number of edges and the number of nodes in the MTG graph, respectively. Alongside these curves, the Bitcoin price is also included for comparison. Since the price of the other four cryptocurrencies is highly correlated with Bitcoin’s price, we only include this coin in the plot. As can be seen, the graph's size is closely related to the price in terms of the number of edges and vertices, and particularly when Bitcoin's price reached $20,000\$$ in late 2017 and a peak appeared in its price chart, we can observe a peak in the MTG size for all five coins. This is due to the fact that as a coin’s price increases, people are more inclined to buy/sell it, resulting in an increased number of generated addresses (i.e. vertices) and transactions (i.e. edges).
\par
For a closer look, we obtained the price of all the coins under review from CoinMarketCap \cite{cap2018cryptocurrency}, and for each coin, we measured the correlation between its price and the size of its MTG graph. Table \ref{tab:pricecorrelation} shows that in all the currencies, there is a positive correlation between the size of the MTG graph and the currency’s price. In particular, this relationship is very strong in Bitcoin, Litecoin and Ethereum.
\par
The plot of the number of edges and the number of nodes in the CMTG graph over time is shown in Fig. \ref{fig:graphsizeofedgescumulative} and \ref{fig:graphsizeofnodescumulative}. As expected, these charts are monotonically increasing due to the cumulative nature of the CMTG graph and shows the growth rate of the CMTG graph of each cryptocurrency over time. In these charts, it can be seen that at some points, the charts intersect, indicating that the number of addresses and transactions of the two currencies is identical and at certain points in time. For example, late in the year 2017 and early in the year 2018, due to the steep rise in the number of addresses of Ethereum, the number of Ethereum addresses exceeded Litecoin and Dash. The reason is that Ethereum has attracted many users by introducing new and unique features such as smart contracts in a short period. The chart also shows that at the beginning of the year 2019, Ethereum has the second largest number of addresses after Bitcoin.
\begin{table}
	\centering
	\caption{Relative growth rate for number of edges and nodes in CMTG graph}
	\begin{tabular}{| c | c | c |}
		\cline{2-3}
		\multicolumn{1}{c|}{} & RGR for edges & RGR for nodes\\
		\hline
		Bitcoin & $0.11206$ & $0.09817$\\ 
		\hline
		Litecoin & $0.07903$ & $0.07092$\\
		\hline
		Ethereum & $0.20173$ & $0.19422$\\
		\hline
		Dash & $0.12264$ & $0.09741$\\ 
		\hline
		Z-Cash & $0.14330$ & $0.12348$\\
		\hline
	\end{tabular}
	\label{tab:relativegrowth}
\end{table}
\par
To be more specific, we calculated the relative growth rate (RGR) for the number of edges and nodes of the CMTG graph from the start of each cryptocurrency until the end of our study, the results of which are presented in Table \ref{tab:relativegrowth}. RGR in the interval $[t_{1} , t_{2}]$ is calculated using the following equation:
\begin{equation}
RGR = \frac{\ln S_{2}  -  \ln  S_{1}}{t_{2} – t_{1}}
\end{equation}
where $S_{i}$ is the size at time $t_{i}$. 
As can be seen, Ethereum has the highest growth rate for both number of edges and number of nodes among the five cryptocurrencies.
\begin{figure}
	\centering
	\begin{subfigure}[t]{0.49\linewidth}
		\includegraphics[width = 1\linewidth]{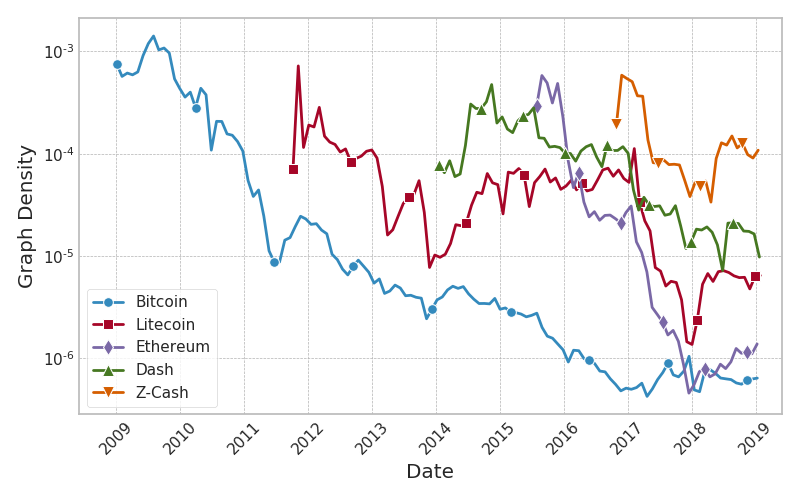}
		\caption{Density of MTG graph}
		\label{fig:graphdensity}
	\end{subfigure}
	\begin{subfigure}[t]{0.49\linewidth}
		\includegraphics[width = 1\linewidth]{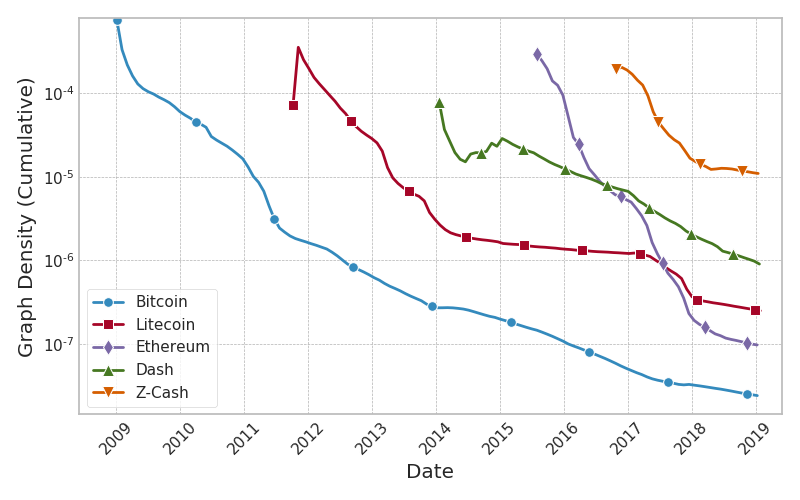}
		\caption{Density of CMTG graph}
		\label{fig:graphdensitycumulative}
	\end{subfigure}
	\caption{Density of transactions graphs over time}
\end{figure}
\begin{table}
	\centering
	\caption{Correlation of price with density of MTG graph}
	\begin{tabular}{| c | c |}
		\cline{2-2}
		\multicolumn{1}{c|}{} & Correlation of price with density\\
		\hline
		Bitcoin & $-0.4981$ \\ 
		\hline
		Litecoin & $-0.5713$ \\
		\hline
		Ethereum & $-0.3312$ \\
		\hline
		Dash & $-0.4787$ \\ 
		\hline
		Z-Cash & $-0.2002$ \\
		\hline
	\end{tabular}
	\label{tab:pricecorrelation2}
\end{table}
\par
Another metric we examined for cryptocurrencies transaction graphs is the density of these graphs. In Fig. \ref{fig:graphdensity} and \ref{fig:graphdensitycumulative}, the density of the MTG and CMTG graphs are plotted over time. As can be seen in Fig. \ref{fig:graphdensity}, the density of the MTG graph is declining in the early years, but rising at the end of the year 2017 and the beginning of the year 2018. This trough created in the density of the MTG graphs coincides with the price peak happened for Bitcoin. To be more precise, for each coin we measured the correlation of price with the density of its MTG graph. Table \ref{tab:pricecorrelation2} shows that for all the currencies, there is a negative correlation between the density of the MTG graph and the price of that currency. This negative correlation is especially strong in Bitcoin, Litecoin, and Dash.
\par
The reason for this can be justified by the fact that to remain anonymous, each user usually generates a new address to enter the cryptocurrency market, and by receiving money from one of the existing addresses, a new edge is created in the transaction graph, thus for each transaction one edge and one node is added to the transaction graph, which results in a linear increase in the number of edges relative to the number of nodes, thereby reducing the graph density. In the early months, due to the steady rise in the price of coins, we are seeing an increase in users willing to invest in the cryptocurrencies market, and as a result, there is an increase in the number of new addresses in these coins, which in turn reduces the density of the transaction graph. But as Bitcoin’s price reached its peak, the decreasing trend of the CMTG density has slowed down or stopped completely. In Fig. \ref{fig:graphdensitycumulative} which illustrates the density of the CMTG graph over time, we observe that the density is most of the time decreasing. The reason might be that the generation of new addresses, which reduces the graph's density, will affect coming months due to the cumulating nature of the graph. But its downward trend has been slower since crossing the Bitcoin’s price peak. The chart also shows that Ethereum’s CMTG density is declining at a much faster rate than other currencies. This implicitly indicates that the number of new Ethereum accounts is increasing sharply, resulting in a decrease in the density.
\begin{figure}
	\centering
	\begin{subfigure}[t]{0.49\linewidth}
		\includegraphics[width = 1\linewidth]{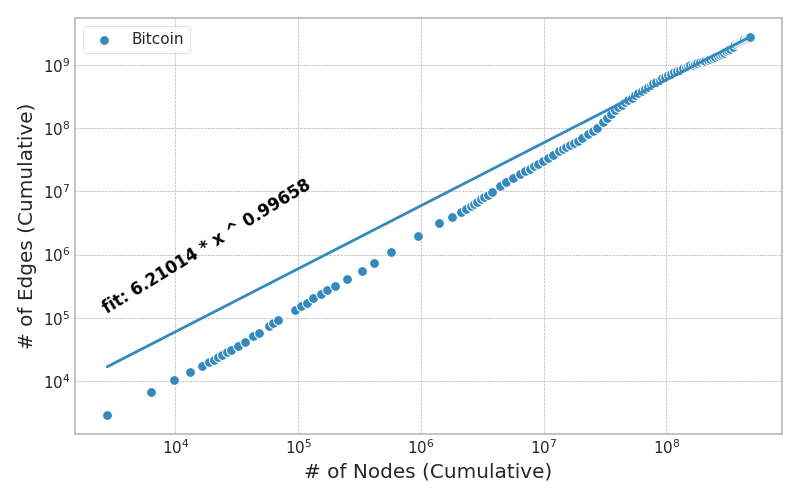}
		\caption{Bitcoin}
		\label{fig:fitbitcoin}
	\end{subfigure}
	\begin{subfigure}[t]{0.49\linewidth}
		\includegraphics[width = 1\linewidth]{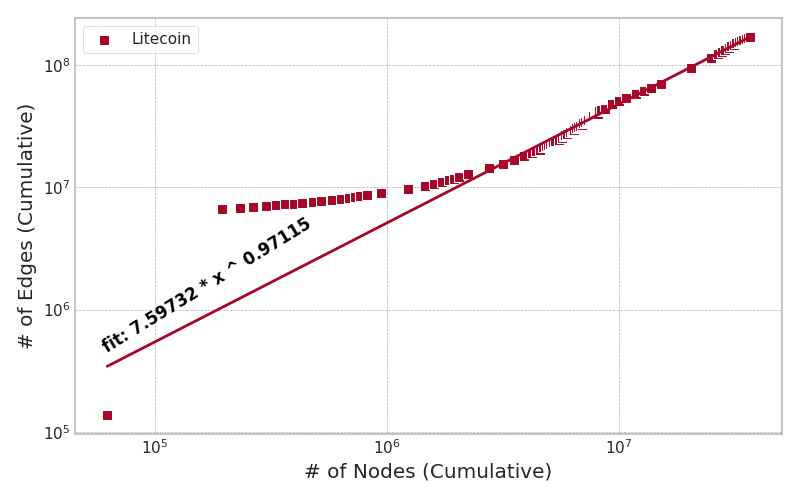}
		\caption{Litecoin}
		\label{fig:fitlitecoin}
	\end{subfigure}
	\begin{subfigure}[t]{0.49\linewidth}
		\includegraphics[width = 1\linewidth]{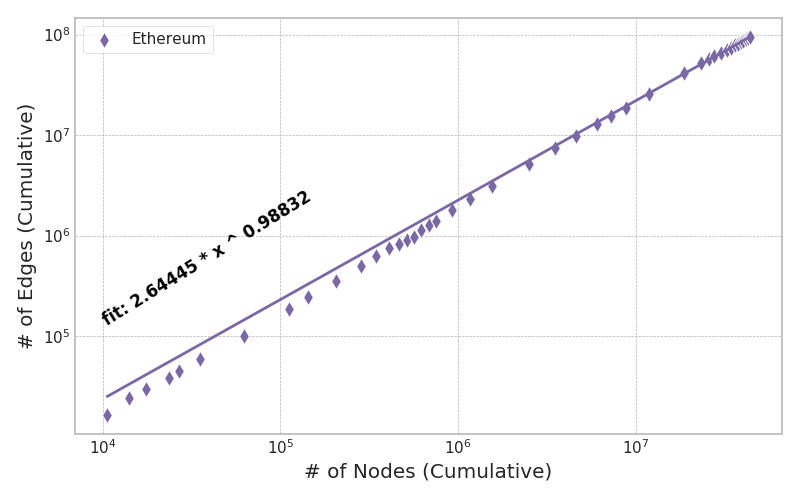}
		\caption{Ethereum}
		\label{fig:fitethereum}
	\end{subfigure}
	\begin{subfigure}[t]{0.49\linewidth}
		\includegraphics[width = 1\linewidth]{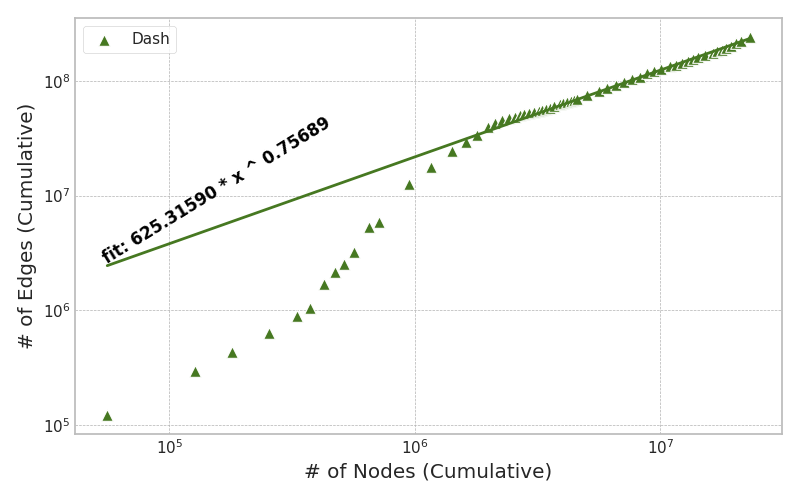}
		\caption{Dash}
		\label{fig:fitdash}
	\end{subfigure}
	\begin{subfigure}[t]{0.49\linewidth}
		\includegraphics[width = 1\linewidth]{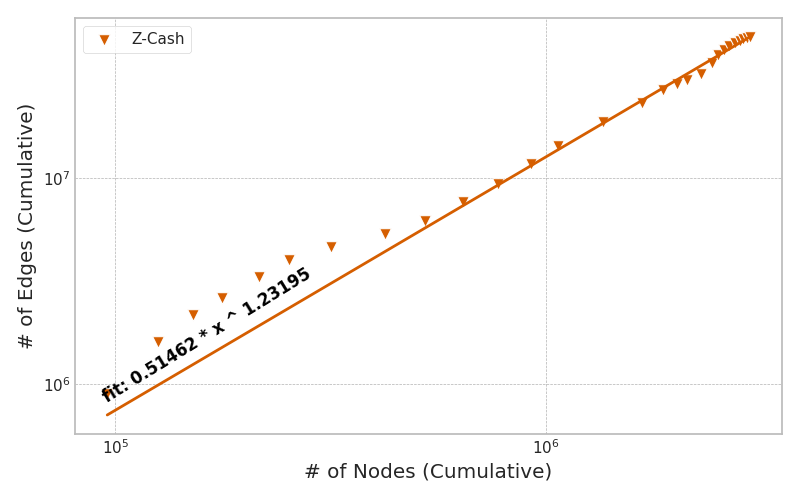}
		\caption{Z-Cash}
		\label{fig:fitzcash}
	\end{subfigure}
	\caption{Fit curve to chart of number of edges in terms of number of nodes in CMTG graph for each coin}
	\label{fig:fitcurve}
\end{figure}
\par
After observing the growth trend of edges and nodes of the CMTG, the next experiment was to determine whether the number of edges of this graph grow linearly with respect to its number of nodes. For this study, we assumed that the number of graph edges is a power function of the number of nodes:
\begin{equation}
n(E) = a \times (n(V)) ^ b
\end{equation}
We estimated the parameters $a$ and $b$ for each of the coins studied. For this purpose, we applied goodness-of-fit tests on the number of edges and the number of nodes in the CMTG graph in different months, we calculated the best curve available by finding the most appropriate $a$ and $b$ for each coin.
\par
As shown in Fig. \ref{fig:fitcurve}, we fit the best possible curve on the data in each of the coins using the power model. The parameters obtained by fitting the curve to the charts are shown in Table \ref{tab:fitcurvetable}, where in all cases the coefficient of determination (Adjusted $R^2$) was over $99\%$. This measure indicates how well points fitted the curve. We observe that in Bitcoin, Litecoin, and Ethereum, the parameter $b$ is very close to 1. Only in the Dash, the value is smaller than 1 and in the Z-Cash, the value is greater than 1, which makes the curve for these two coins slightly distant from the linear curve. 
\begin{table}
	\centering
	\caption{Estimated parameters for the best-fitted curve on the chart of the number of edges by the number of nodes in CMTG graph}
	\begin{tabular}{| c | c | c | c |}
		\cline{2-4}
		\multicolumn{1}{c|}{} & $a$ & $b$ & $Adjusted\;R^2$ \\
		\hline
		Bitcoin & $6.21$ & $0.996$ & $99.56\%$ \\ 
		\hline
		Litecoin & $7.60$ & $0.971$ & $99.67\%$ \\
		\hline
		Ethereum & $2.64$ & $0.988$ & $99.98\%$ \\
		\hline
		Dash & $625.32$ & $0.757$ & $99.44\%$ \\ 
		\hline
		Z-Cash & $0.51$ & $1.232$ & $99.53\%$ \\
		\hline
	\end{tabular}
	\label{tab:fitcurvetable}
\end{table}

\begin{figure}
	\centering
	\begin{subfigure}[t]{0.49\linewidth}
		\includegraphics[width = 1\linewidth]{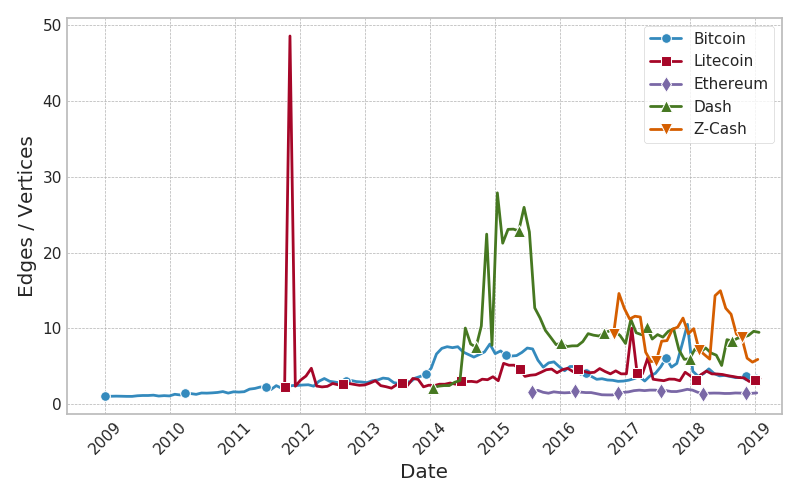}
		\caption{Timeseries of edge-to-vertex ratio for MTG}
		\label{fig:edgestovertices}
	\end{subfigure}
	\begin{subfigure}[t]{0.49\linewidth}
		\includegraphics[width = 1\linewidth]{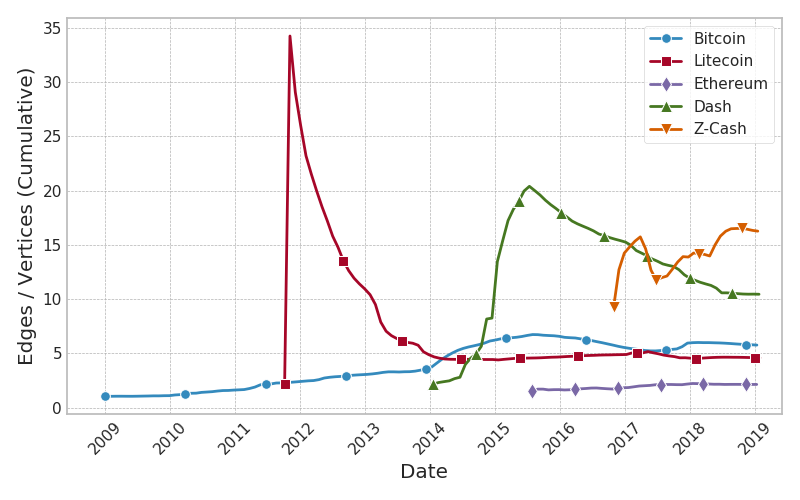}
		\caption{Timeseries of edge-to-vertex ratio for CMTG}
		\label{fig:edgestoverticescumulative}
	\end{subfigure}
	\caption{Edge-to-vertex ratio of transaction graphs}
\end{figure}
\par
Since the linear increase in the number of edges relative to the number of nodes reduces the graph density, another parameter examined on MTG and CMTG graphs is the ratio of the number of edges to the number of nodes. As shown, in most cryptocurrencies, the ratio of the number of edges to the number of nodes is linear or close to linear. The ratio of the number of edges to the number of nodes indicates the slope of this linear relation. In Fig. \ref{fig:edgestovertices} the time series of the edge-to-node ratio of the MTG graph is plotted. As you can see, except for some specific cases on the chart, in most coins this ratio fluctuates between 1 and 15, but its overall trend is steady. In Dash, in a time interval, this value is about 25 to 30, but then in the following interval, its trend is fixed below 10. But in Litecoin, in the second month of its creation, a noticeable phenomenon has happened and this ratio is increased by up to 40, and returns to normal in the following months, and its trend is almost constant. Further investigation into the Litecoin’s blockchain revealed that a sequence of transactions with specific patterns occurred in mid-November 2011, which generated a large number of edges in the transaction graph.
\begin{table}
	\centering
	\caption{Number of edges and nodes of Litecoin's MTG graph in its first 3 months}
	\begin{tabular}{ | c | c | c |}
		\cline{2-3}
		\multicolumn{1}{c|}{} & Number of nodes & Number of edges \\
		\hline
		First month & $62,321$ & $137,825$ \\ 
		\hline
		Second month & $134,607$ & $6,538,828$ \\
		\hline
		Third month & $40,981$ & $96,812$ \\
		\hline
	\end{tabular}
	\label{tab:litecoinmtgstats}
\end{table}
\par
As shown in Table \ref{tab:litecoinmtgstats}, the number of nodes in the second month of MTG graph is about 2 to 3 times its prior and subsequent months, but its number of edges is about 50 to 70 times its prior and subsequent months.
\begin{figure}
	\centering
	\begin{subfigure}[t]{0.3\linewidth}
		\includegraphics[width = 1\linewidth]{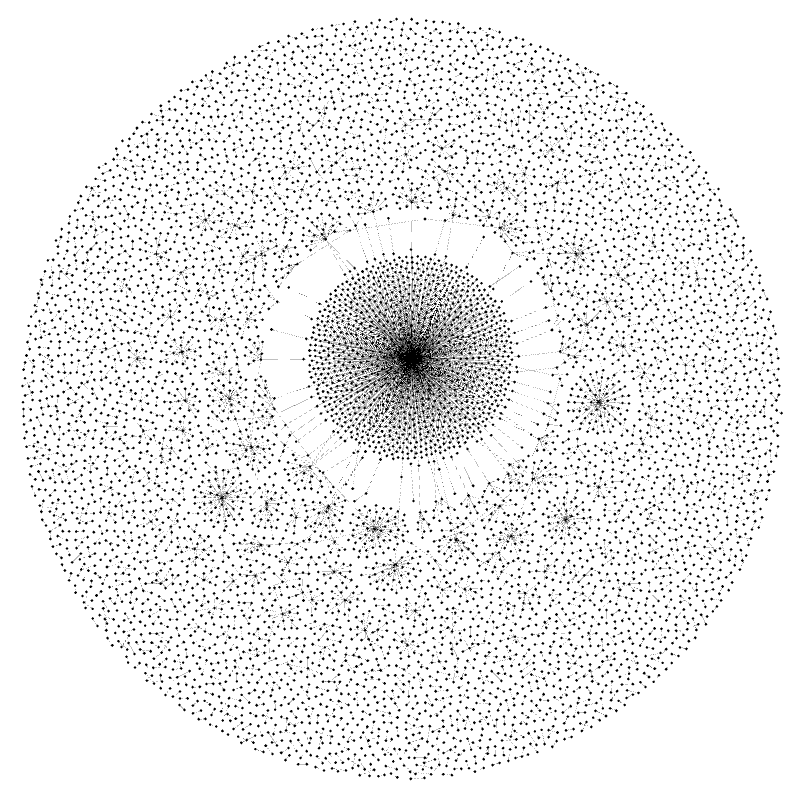}
		\caption{$MTG_0$ graph (First month)}
		\label{fig:litecoinmtg0}
	\end{subfigure}
	\begin{subfigure}[t]{0.3\linewidth}
		\includegraphics[width = 1\linewidth]{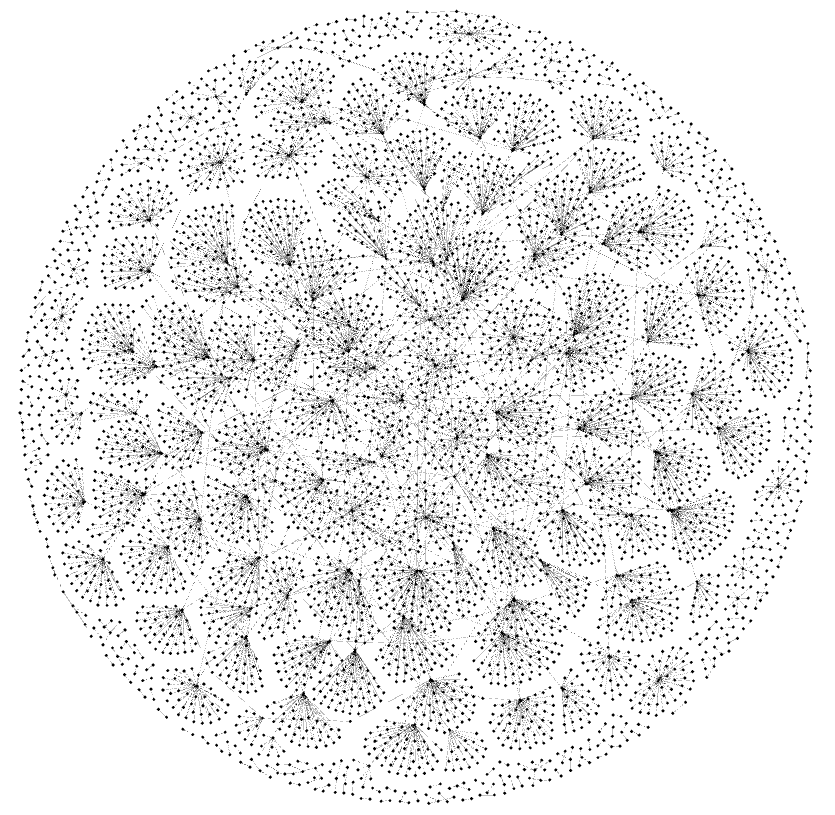}
		\caption{$MTG_1$ graph (Second month)}
		\label{fig:litecoinmtg1}
	\end{subfigure}
	\begin{subfigure}[t]{0.3\linewidth}
		\includegraphics[width = 1\linewidth]{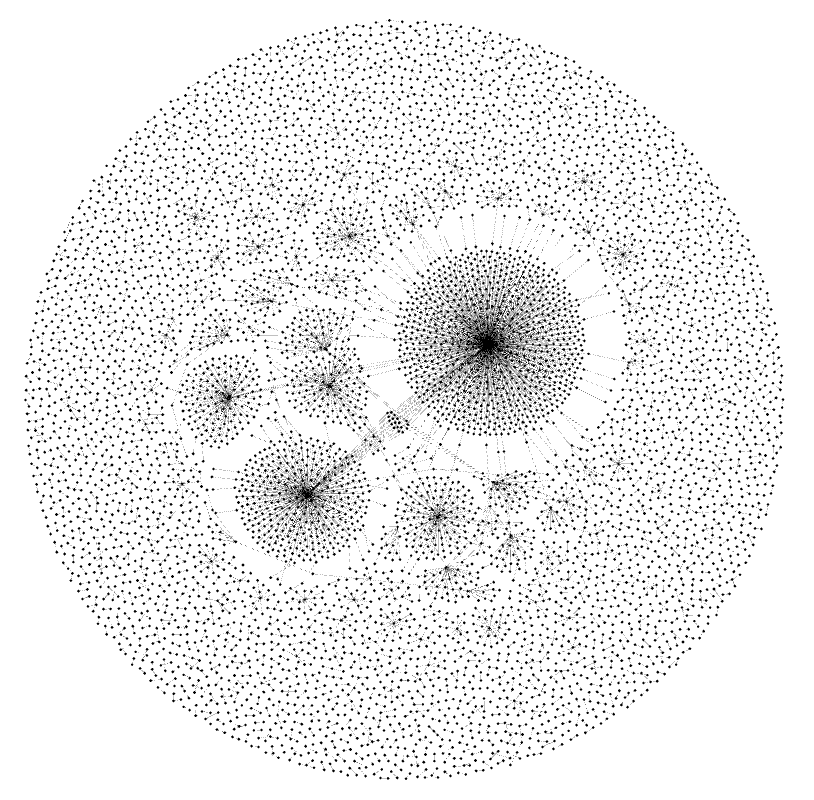}
		\caption{$MTG_2$ graph (Third month)}
		\label{fig:litecoinmtg2}
	\end{subfigure}
	\caption{Visualization of Litecoin's MTG graph in its first 3 months}
	\label{fig:litecoinmtg}
\end{figure}
\par
Fig. \ref{fig:litecoinmtg} shows an overview of the first three months of Litecoin MTG graphs. Due to the large size of the graphs, their shape is plotted by taking random samples of approximately 5000 edges from each graph. As can be seen in Fig. \ref{fig:litecoinmtg0} and \ref{fig:litecoinmtg2}, in the first and third months we see the normal structure of the transaction graph, which behaves similarly to social networks. In other words, the network has a very limited number of high degree hubs and many low degree nodes that are connected to these hubs. But in the second month, when there is a dramatic increase in the number of edges and vertices, we observe a lot of hubs of similar size that are interconnected, and this pattern is because of specific transactions that took place in the second month.
\begin{figure}
	\centering
	\includegraphics[width = 0.49\linewidth]{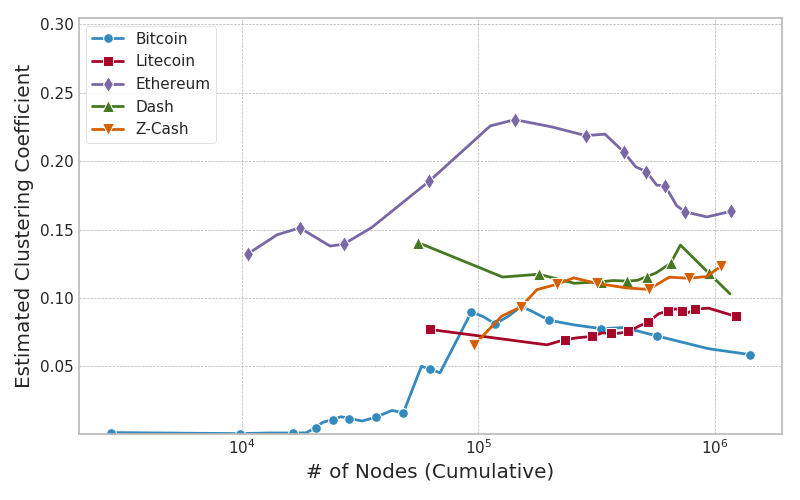}
	\caption{Clustering coefficient of CMTG vs. the number of its nodes}
	\label{fig:clusteringcoefficientcumulative}
\end{figure}
\par
Another calculation performed on the CMTG graph was the clustering coefficient. The clustering coefficient is used as a measure of the degree of the graph node's willingness to create a cluster. Given that it is very difficult to perform these computations on large graphs, we performed the computations only in the months when the CMTG graph were less than one million nodes, and plot the charts in terms of the number of nodes and use the estimated method to calculate the clustering coefficient. Fig. \ref{fig:clusteringcoefficientcumulative} illustrates the clustering coefficient of CMTG. We have also calculated the correlation coefficient between the clustering coefficient and the number of CMTG nodes (see Table~\ref{tab:cliqueclustering}). As shown in Table \ref{tab:cliqueclustering}, there is a positive correlation between the number of CMTG graph nodes and its clustering coefficient for all coins.
\begin{figure}
	\centering
	\includegraphics[width = 0.49\linewidth]{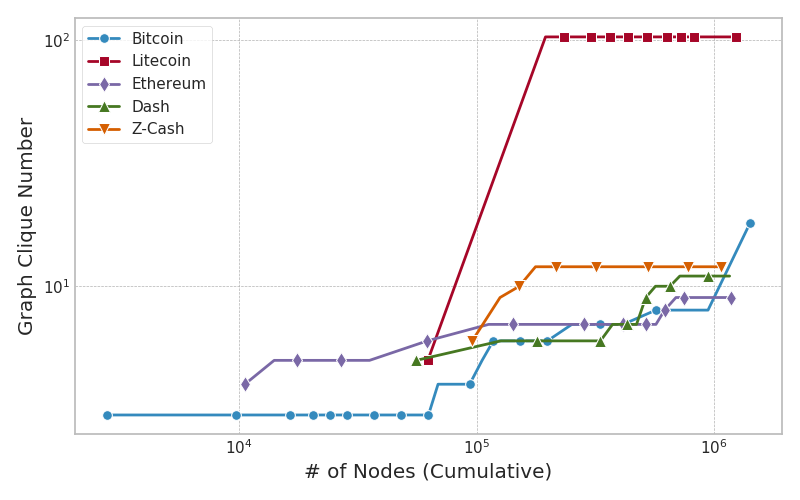}
	\caption{Maximum clique size of CMTG vs. the number of its nodes}
	\label{fig:maxcliquecumulative}
\end{figure}
\par
We also calculated the maximum clique size for the CMTG. In the transaction graph, a maximum clique represents the highest number of accounts in which there was a transaction between each pair of them. The presence of large cliques usually indicates a very strong relationship between the accounts in the clique. Finding the maximum clique size is an NP-hard problem, and doing so on large graphs are computationally expensive, thus we found the maximum clique only for the months when the CMTG had less than one million nodes. Fig. \ref{fig:maxcliquecumulative} shows the maximum clique size versus the number of nodes in the CMTG of the five cryptocurrencies. Since in cumulative transaction graph, we are only adding new nodes and edges with time, the maximum clique size is always increasing. In Fig. \ref{fig:maxcliquecumulative} we observe that the maximum clique of Litecoin’s CMTG after the first month is much larger compared to other cryptocurrencies. This is because in the second month of Litecoin, a maximum clique of size 104 is observed, and in subsequent months no larger maximum clique is found. To examine more precisely the relationship between maximum clique size and the number of CMTG nodes, we calculated the correlation between these two variables for the five coins, the results of which are presented in Table \ref{tab:cliqueclustering}. According to these results, there are relatively strong relationships between the maximum clique size and the number of nodes in all coins’ CMTGs.
\begin{table}
	\centering
	\caption{Correlation of maximum clique size and clustering coefficient with the number of nodes in CMTG}
	\begin{tabular}{| c | c | c |}
		\cline{2-3}
		\multicolumn{1}{c|}{} & clustering coefficient & maximum clique size \\
		\hline
		$\abs{V(\text{Bitcoin CMTG})}$ & $0.51848$ & $0.63372$ \\ 
		\hline
		$\abs{V(\text{Litecoin CMTG})}$ & $0.69864$ & $0.43774$ \\
		\hline
		$\abs{V(\text{Ethereum CMTG})}$ & $0.13774$ & $0.69444$ \\
		\hline
		$\abs{V(\text{Dash CMTG})}$ & $0.07195$ & $0.93368$ \\ 
		\hline
		$\abs{V(\text{Z-Cash CMTG})}$ & $0.62751$ & $0.52585$ \\
		\hline
	\end{tabular}
	\label{tab:cliqueclustering}
\end{table}
\par
Users of cryptocurrencies are constantly changing and expanding. The price of these currencies is also very volatile. As the prices of these currencies rise, new customers' desire to buy cryptocurrencies increases. Also, at a downward trend for these coins’ price, some people prefer to sell their coins and convert them into more stable assets, such as precious metals and fiat currencies, to maintain their value. One of the studies we performed on transaction graphs was the analysis of the repetition ratio of nodes and edges in each month compared to its previous month. In other words, we examined what percentage of accounts that have made a transaction in a month, had also transacted in the previous month, and what percentage of the transactions, that took place this month, had also occurred between the same addresses in the previous month. The closer these metrics are to the 1, the number of new accounts and new financial transactions in the network is smaller.
\begin{figure}
	\centering
	\begin{subfigure}[t]{0.49\linewidth}
		\includegraphics[width = 1\linewidth]{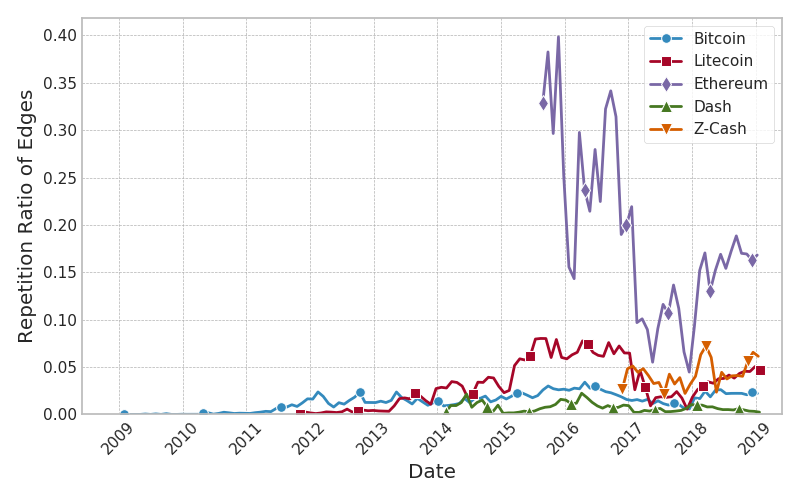}
		\caption{Repetition ratio of edges}
		\label{fig:repetitionratio}
	\end{subfigure}
	\begin{subfigure}[t]{0.49\linewidth}
		\includegraphics[width = 1\linewidth]{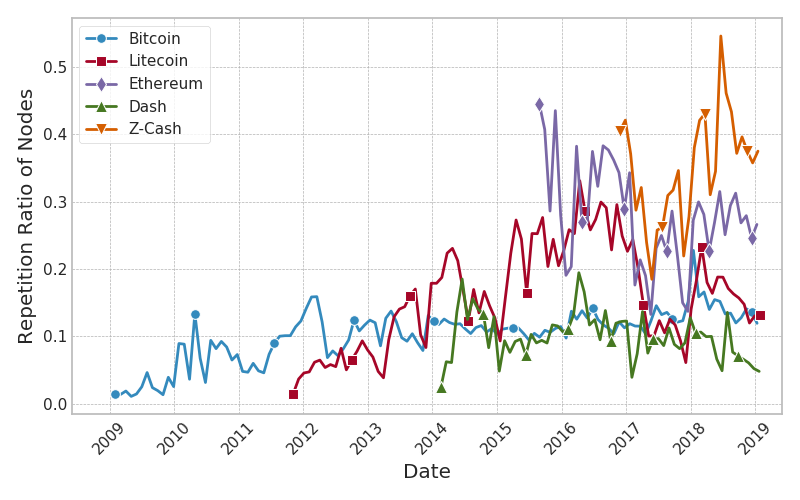}
		\caption{Repetition ratio of nodes}
		\label{fig:repetitionratiocumulative}
	\end{subfigure}
	\caption{Repetition ratio of MTG’s nodes and edges over time}
\end{figure}
\par
Given that in the CMTG graph, because of its cumulative, most nodes and edges between them are duplicate compared to the previous month, this parameter tends to value 1. So calculating this parameter on the CMTG graph makes no sense. So we did the calculations just on the MTG graph. In Figures \ref{fig:repetitionratio} and \ref{fig:repetitionratiocumulative}, time-series of the edge and node repetition ratio are plotted for the MTG. These curves have a lot of fluctuations, but it is observable that Dash and Bitcoin have the lowest repetition ratio, and the highest values of this metric belong to Z-Cash and Ethereum, and Litecoin is in the middle. The node repetition ratio for Z-Cash has reached above 0.5 at some points, which is a quite large value for this metric. The common reason for creating a new address for each transaction in cryptocurrencies is to prevent finding a connection between different addresses of the same user. But in Z-Cash, due to the use of zero-knowledge proofs, it is possible to hide the sender, the receiver and the amount of money in a transaction, and as a result, users do not need to create a new address to remain anonymous. Also in the Ethereum, because of its account-based blockchain and the singularity of the sender and recipient of each transaction, we observe a high repetition ratio for both nodes and edges.
\begin{figure}
	\centering
	\includegraphics[width = 0.49\linewidth]{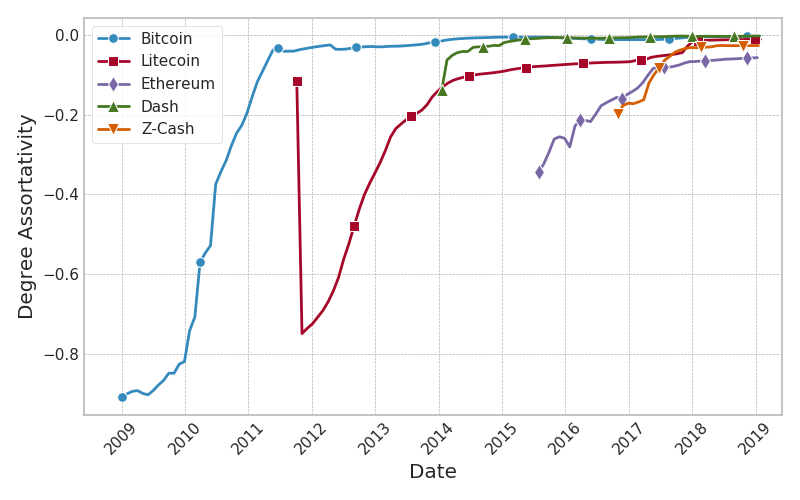}
	\caption{Degree assortativity in CMTG graph}
	\label{fig:degreeassortativity}
\end{figure}
\par
One of the important metrics usually calculated on social networks is assortativity. This property shows whether nodes tend to communicate with other nodes which are similar to them or not. There can also be a situation where nodes are neutral in communicating with other nodes. In this study we calculated degree assortativity for the transaction graph. In other words, we want to see if accounts were more likely to have a transaction with their counterparts or vice versa. The degree in the transaction graph for a node is the number of distinct accounts with which it has transacted. Fig. \ref{fig:degreeassortativity} shows the degree assortativity of the CMTG over time. As can be seen, for all the coins at the beginning of their activity, the assortativity is negative. The reason for this is that we have considered a supernode for coinbase that gives the miners the rewards for block generation. As a result, for each coin, we have created a hub at the beginning, and most of the early users are connected to this high-degree supernode. With expansion of the cryptocurrency network and growing number of its users, this metric tends to zero for all coins, because most of the transaction are done between ordinary users. As a result, we can say that the transaction graph is a non-assortative graph, i.e. the accounts keep the balance between transacting with their similar accounts and those larger or smaller than them. Another noticeable case in the degree assortativity plot is the sharp decrease of this metric in the second month of Litecoin’s lifetime. As discussed above, in the second month of Litecoin, a large number of hubs of the same size have been created, each of them connected to several small nodes, mostly of degree $1$ (see Fig. \ref{fig:litecoinmtg1}). Since most of the transaction graphs include these centers with their low-degree neighbors, we see a dramatic decrease in the degree assortativity of Litecoin’s CMTG, making it an extremely disassortative graph. 
\begin{figure}
	\centering
	\begin{subfigure}[t]{0.49\linewidth}
		\includegraphics[width = 1\linewidth]{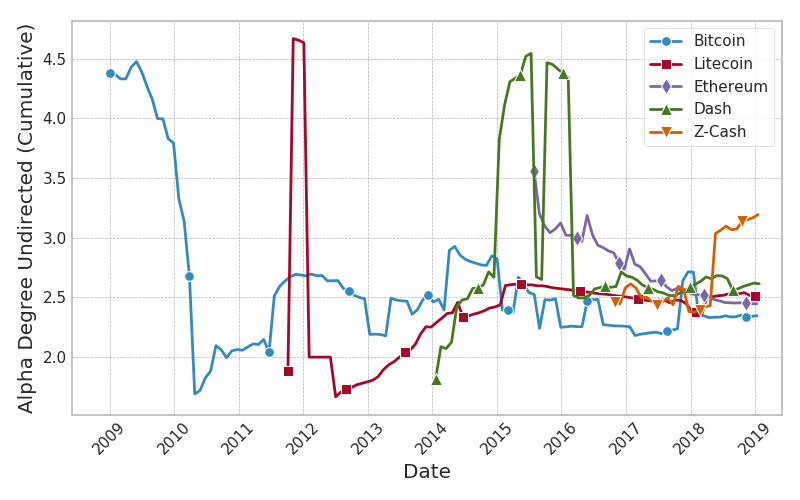}
		\caption{Degree of undirected graph}
		\label{fig:alphadegreeundirected}
	\end{subfigure}
	\begin{subfigure}[t]{0.49\linewidth}
		\includegraphics[width = 1\linewidth]{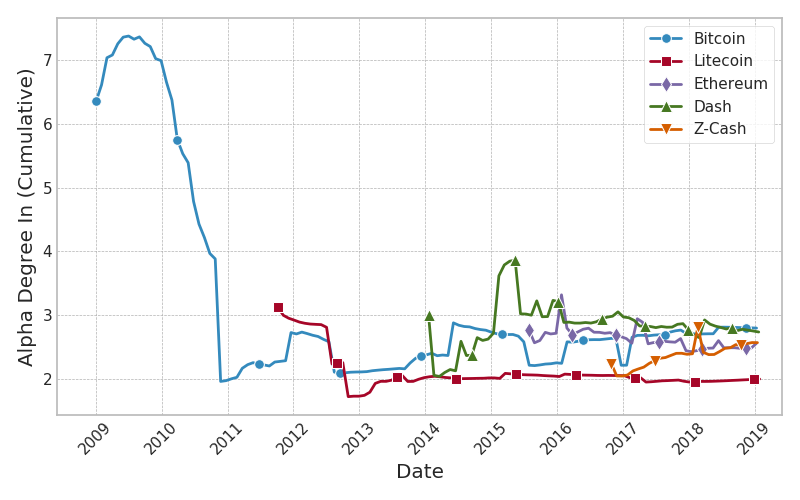}
		\caption{In-degree of directed graph}
		\label{fig:alphadegreein}
	\end{subfigure}
	\begin{subfigure}[t]{0.49\linewidth}
		\includegraphics[width = 1\linewidth]{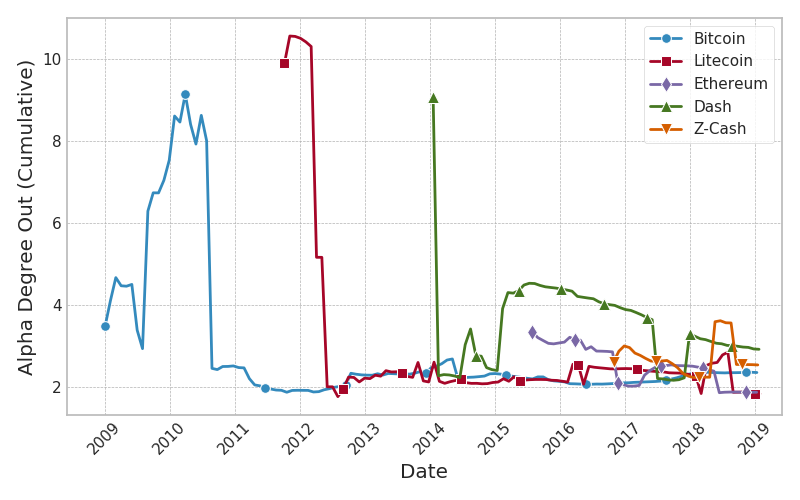}
		\caption{Out-degree of directed graph}
		\label{fig:alphadegreeout}
	\end{subfigure}
	\begin{subfigure}[t]{0.49\linewidth}
		\includegraphics[width = 1\linewidth]{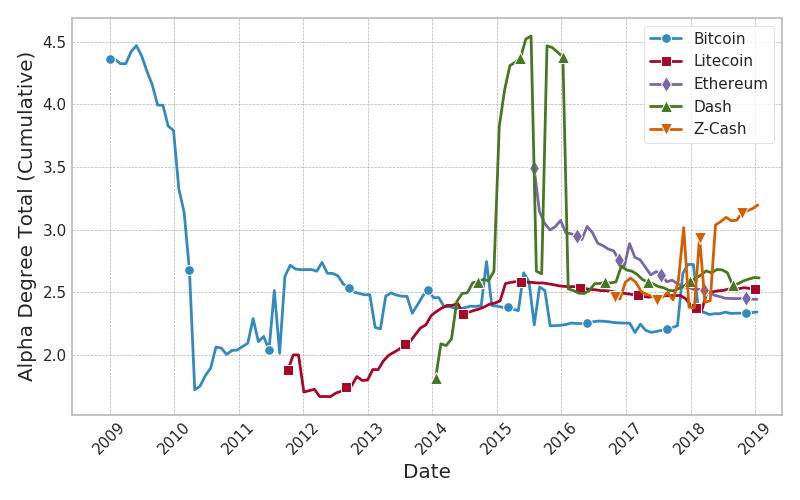}
		\caption{Total-degree of directed graph}
		\label{fig:alphadegreetotal}
	\end{subfigure}
	\caption{Alpha parameter of the Power-Law distribution over the degrees sequence of the CMTG}
	\label{fig:alpha}
\end{figure}
\par
Understanding the overall structure of a cryptocurrency’s transaction graph gives us a comprehensive view of the nature of that currency. One way to figure out how money is distributed among accounts is to find the sequence of nodes’ degrees in the graph. It is shown that in most social networks, the degree sequence follows a Power-Law distribution \cite{barabasi412}. In the cryptocurrencies transaction graph, this logic is also expected to be dominant. For this purpose, in the CMTG, we calculated the degree sequence of vertices for each month, and by fitting a Power-Law distribution to this sequence, we computed the main parameter of this distribution ($\alpha$). The Power-Law distribution is defined as follows:
\begin{equation}
\begin{aligned}
& f(x) = C x ^ {-\alpha} & \mbox{  for  } x \geq x_{min},
\end{aligned}
\end{equation}
where $C = (\alpha-1)(x_{min})^{\alpha-1}$ is a constant. 
To calculate the degree sequence, we considered both directed and undirected CMTG and calculated the degree sequence for each separately. For the directed graph, we calculated three sequences of in-degree, out-degree, and the total degree. Note that the total degree of a node in the directed graph might be different from the degree of that node in the undirected graph, because if there is a transaction from $v_1$ to $v_2$, and another transaction from $v_2$ to $v_1$, in the directed version we consider them as two different edges, but in the undirected graph there is only one edge between $v_1$ and $v_2$. We then fitted a separate Power-Law distribution on each of these four degree sequences and calculated its parameter $\alpha$ for the best-fitted distribution and plotted this parameter variation over time for each coin in Fig. \ref{fig:alpha}.
\par
We can see that in all cases the parameter $\alpha$, after some fluctuations, converges to a steady state, usually between $2$ and $3$, which shows that we are dealing with a scale-free network \cite{BARABASI2001559}. We know that power law is a top heavy (extremely right skewed) distribution, and as $\alpha \rightarrow 2$ this asymmetry gets increasingly more extreme, with a smaller and smaller fraction of the nodes holding a greater and greater proportion of the edges in the network.
\par
Also in the second month of Litecoin’s lifetime, that we discussed earlier, we see a jump in Fig. \ref{fig:alphadegreeundirected} and \ref{fig:alphadegreeout}, but this jump is not seen in Fig. \ref{fig:alphadegreein} and \ref{fig:alphadegreetotal}. This is due to an anomaly in the second month of Litecoin which in that a large number of hubs have been created and it caused a small tail in the power-law distribution. We see this jump only in the figures related to the undirected degree and the out-degree distribution and we can infer that the edges connected to these hubs are mostly incoming edges, because the larger values of $\alpha$ for out-degree distribution happen when the out-degrees of the hubs are similar to ordinary nodes of the network, i.e. the hubs’ in-degrees are large. 
\section{Conclusion}
In this paper, we studied various structural properties of the transaction graphs of Bitcoin, Litecoin, Ethereum, Dash, and Z-Cash. We also examined how the cryptocurrencies transaction graphs evolve over time and discussed their dynamics. In this study, we found that in all the coins, the number of edges of cumulative transaction graphs increase linearly with respect to the number of nodes. We observed that in the cumulative transaction graph, the density is always decreasing. We also investigated the relationship between the transaction graph size with maximum clique size and clustering coefficient, and observed that there is a positive correlation between these metrics. Also, the number of nodes and edges per month of the non-cumulative transaction graph is closely related to the price of these currencies. In the second month of Litecoin, we found and described a set of abnormal transactions and examined their effect on some graph metrics. We also showed that the edge and node repetition ratio in non-cumulative transaction graphs of Ethereum and Z-cash is higher in comparison with other currencies, due to their specific technical properties. Another important finding of this study is that the transaction graphs of all examined coins are becoming non-assortative as they grow larger over time.

\bibliographystyle{elsarticle-num}  


\end{document}